\documentclass[%
 reprint,
 amsmath,amssymb,
 aps,
floatfix,
]{revtex4-2}

\usepackage{graphicx}% Include figure files
\usepackage{dcolumn}% Align table columns on decimal point
\usepackage{bm}% bold math
\usepackage{xcolor}
\usepackage{textgreek}
\begin{document}

\preprint{APS/123-QED}

\title{Quadristor: a novel device for superconducting electronics}

\author{Sara Chahid}
\affiliation{Advanced Physics Laboratory, Institute for Quantum Studies, Chapman University, Burtonsville, MD 20866, USA}

\author{Serafim Teknowijoyo}
\affiliation{Advanced Physics Laboratory, Institute for Quantum Studies, Chapman University, Burtonsville, MD 20866, USA}

\author{Armen Gulian}
\email[Corresponding author: ]{gulian@chapman.edu}
\affiliation{Advanced Physics Laboratory, Institute for Quantum Studies, Chapman University, Burtonsville, MD 20866, USA}
% \altaffiliation[Also at ]{Physics Department, XYZ University.}%Lines break automatically or can be % forced with \\

\date{\today}% It is always \today, today,

\begin{abstract}

We designed and experimentally demonstrated a four-terminal superconducting
device which can function as a non-latching (reversible) superconducting switch 
from the diode regime to the
resistive state by applying a control current much smaller than the main
transport current. The device utilizes a vortex-based superconducting diode
mechanism which is switched back and forth via the injection of flux quanta
through auxiliary current leads. Various applications in superconducting
electronics can be foreseen.

\end{abstract}

\maketitle

%\tableofcontents

%\section{Introduction}

Simultaneous breaking of time reversal symmetry and inversion symmetry
generates the superconducting diode effect  \cite{Tokura18,Wakatsuki18,Hoshino18,Ando20,Ideue20,BaumgartnerNN,Wu22,Strambini22}. 
The time reversal symmetry could
be broken via an externally applied magnetic field or internal inclusions of
magnetic micro-clusters, while the inversion symmetry could be broken in several ways:
\cite{Yuan22,Ilic22,Daido22,Karabassov22,He22,Ando20,BaumgartnerNN,Bauriedl22,Wakatsuki17,Shin21,BaumgartnerIOP,Hou22,Suri22,Hope21,Vodolazov05,Chahid22}.

The recent spike of research in the area of superconducting diodes is paving
a road towards the future practical application of these novel devices in
superconducting electronics. Moreover, it inspires substantiated hopes that
the next stages of work, in analogy with semiconductor electronics, will be
accomplished by developing superconducting transistors. By a ``transistor" we
refer to a regulated diode, whose resistive state can be controlled by an
externally applied signal exerting much less power than the one it controls.
In this report, we introduce such a device. Because of the specifics of
superconductivity, where the devices are current-biased, the controlling
agent in our design is a current supplied by two auxiliary leads. This
justifies the term ``quadristor": two leads for the transport and two leads 
for the control currents. Via this relatively small control current,
the diode effect can be turned off and on by demand without noticeable
latching effect and can function as a fast switch or as a signal
controller/amplifier.

We used $\mathrm{Nb}_{3}\mathrm{Sn}$ films on sapphire substrates with $%
T_{c}>17$ $\mathrm{K}$, Fig.~1. 
\begin{figure}
    \centering
    \includegraphics[width=\linewidth]{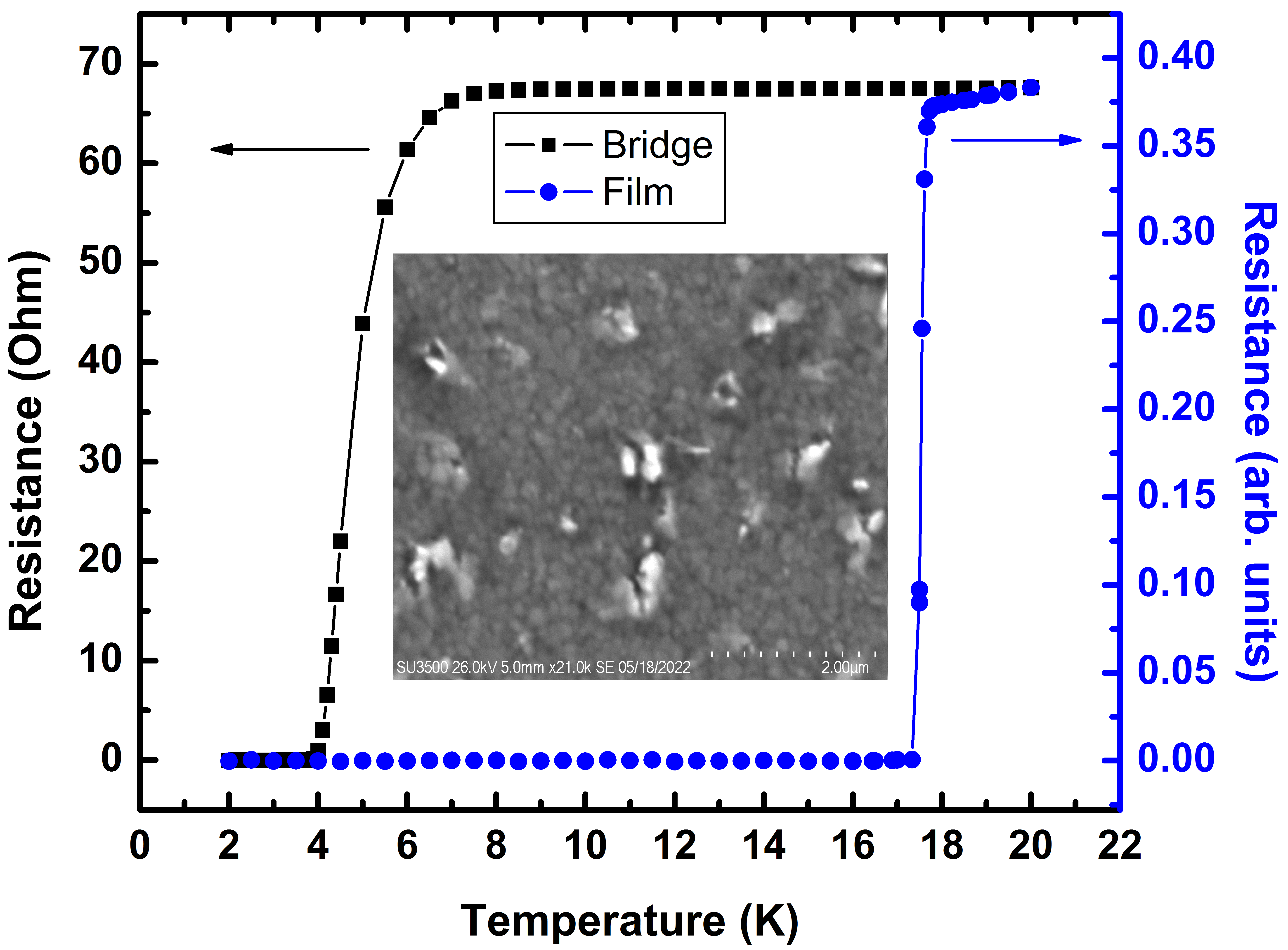}
    \caption{Superconducting resistive
transition in our $\mathrm{Nb}_{3}\mathrm{Sn}$ film
and the processed bridge from it. Resistive transition of the bridge shows reduction of $T_c$ (in this work, we did not attempt its restoration to 17 K by annealing as we did in our previous work \cite{Chahid22} and used the bridge as prepared). SEM image (inset)
shows the material's granularity with the average grain size about 200 nm, 
as well as larger in size nanomountains which, as the EDX revealed, contain excess amount of Sn (further
information about the latter could be obtained in \cite{Chahid22}).}
    \label{fig1}
\end{figure}

The device was fabricated by a multistage lithographic patterning process
described in Fig.~2 (the details of deposition and lithography patterning
are described elsewhere \cite{Chahid22}).

\begin{figure}[h!]
    \centering
    \includegraphics[width=\linewidth]{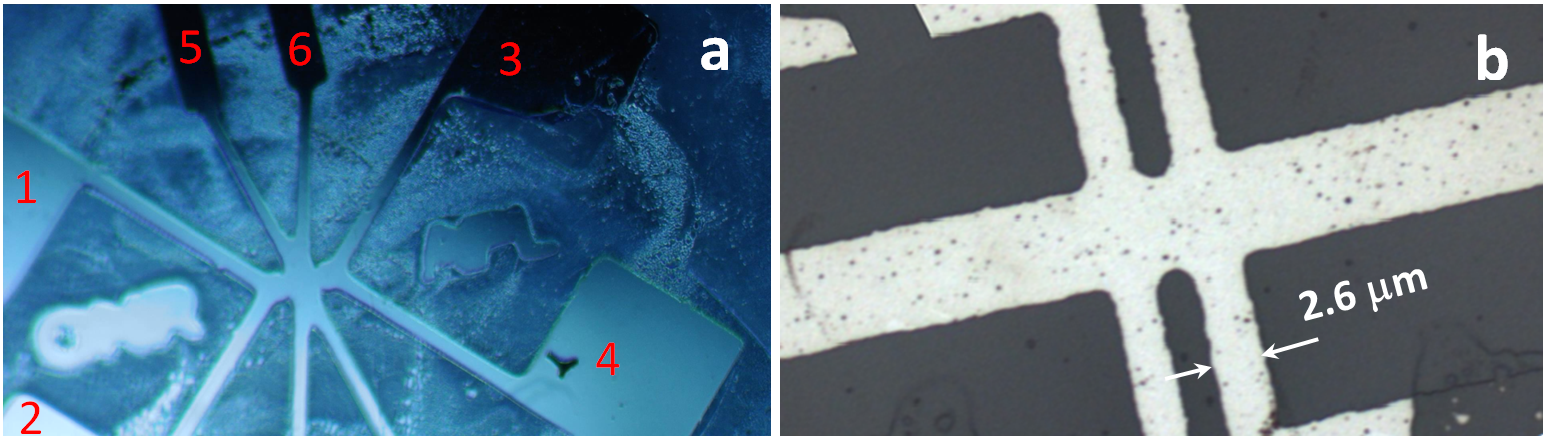}
    \caption{Consecutive steps in lithographic processing of the device. (a)
Larger scale pattern obtained after 3D-printing and ion milling; numeric
labels indicate contact pads (two more of them, $7$ and $8$ are not
shown); the lateral size of the bridge is $\approx 100$ $\mathrm{\mu m}$.
(b) Final layout of the central part of the bridge after projection
photolithography and ion milling. The projection mask had nanoholes which
transferred onto the $\mathrm{Nb}_{3}\mathrm{Sn}$ material. Their role will
be discussed further in the text.}
    \label{fig2}
\end{figure}

Ion milling affects the physical properties of bridges and reduces the
critical temperatures (see. e.g., \cite{Suri22,Chahid22}). In the current case, the
bridge had $T_{c}\sim 4$ $\mathrm{K}$ (Fig.~1).

\begin{figure}
    \centering
    \includegraphics[width=\linewidth]{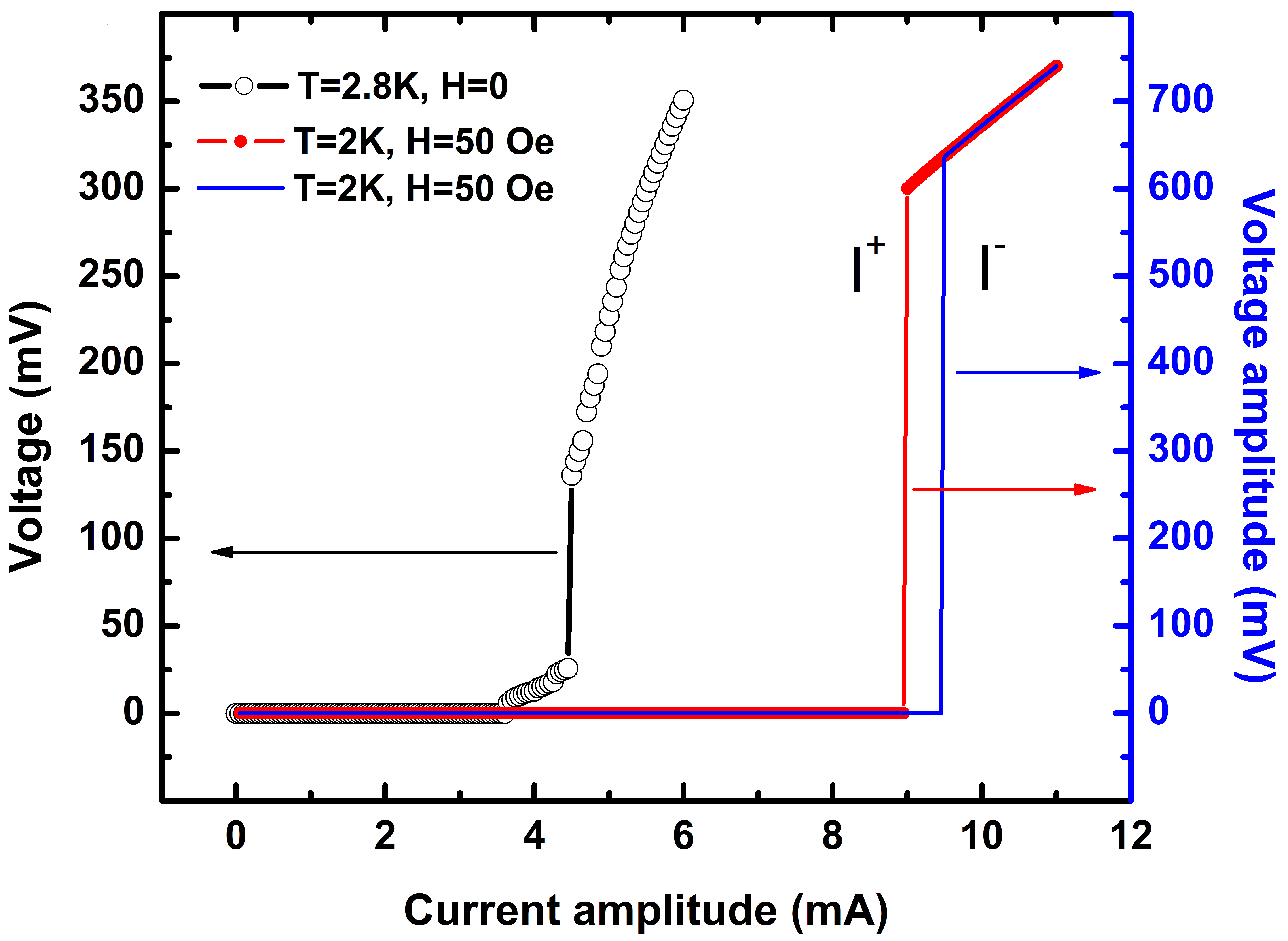}
    \caption{Voltage-current curve 
    (measured by Quantum Design PPMS system) at $2.8$ \textrm{K} and $H=0$
demonstrates steps caused by certain re-structuring of vortex patterns
in the bridge \cite{Suri22,Berdiyorov09,Lotero14,Vodolazov07,Chahid22}.
Voltage-current characteristics 
(measured by Keithley 6221 current source and 2182A nanovoltmeter)
at $T=2$ \textrm{K} in the PPMS cryostat at $H=50$ \textrm{Oe} 
(at this and all other measurements reported in this article, the direction 
of \textbf{H} is orthogonal to the surface of the film)
at positive and negative directions of the dc current flow.}
    \label{fig3}
\end{figure}

As was reported previously \cite{Chahid22}, the bridges prepared in this manner
demonstrate a noticeable critical current difference $\Delta I=I^{+}-I^{-}$.
For the bridge under study, we reached $\Delta I=0.5$ $\mathrm{mA}$ at $T=2$ 
$\mathrm{K}$ and $H=50$ \textrm{Oe} (Fig.~3; similarly with \cite{Suri22,Chahid22}, at
positive values of applied magnetic field, $I^{+}<I^{-}$ and vice versa)
with very robust diode effect (Fig.~4).

\begin{figure}
    \centering
    \includegraphics[width=\linewidth]{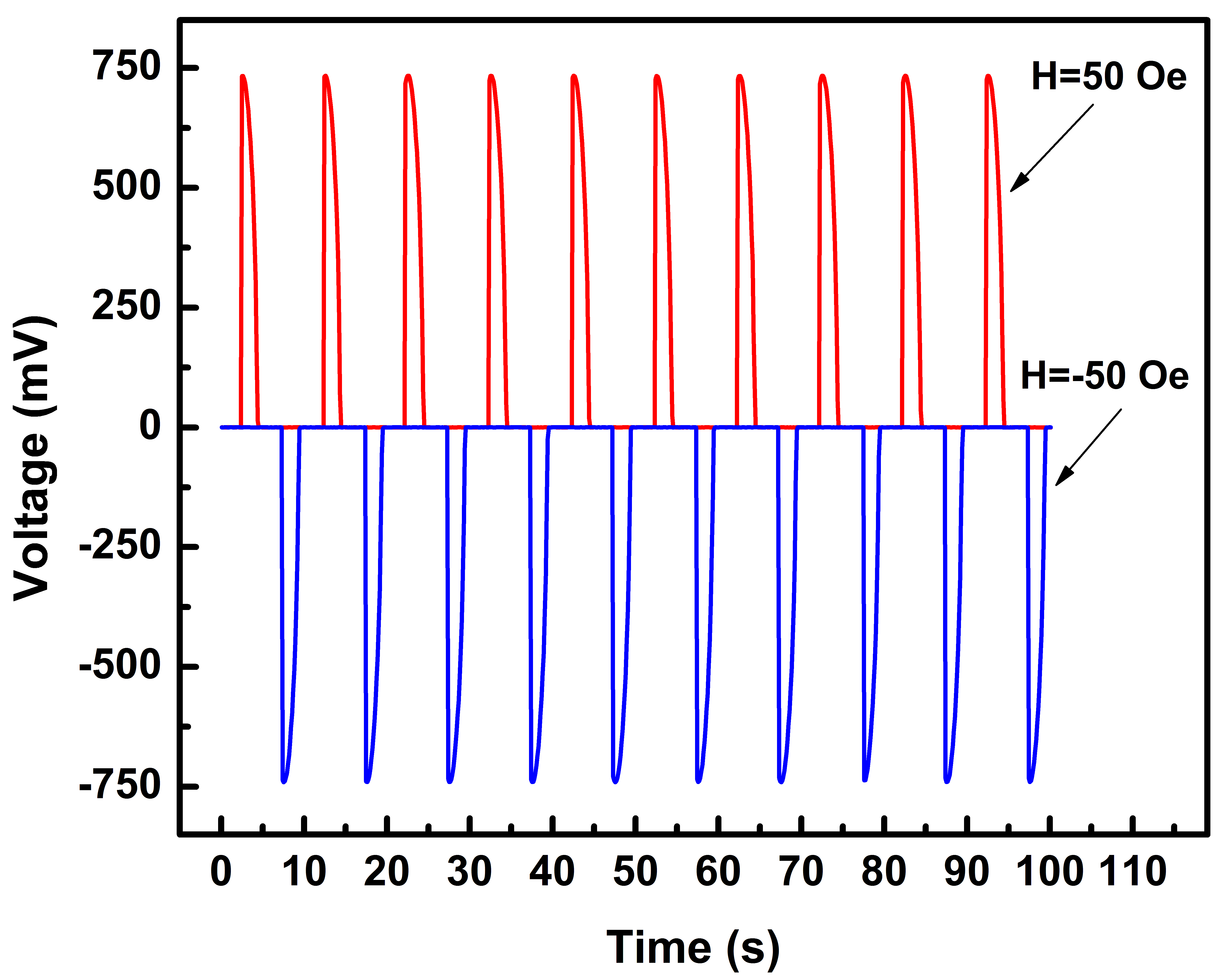}
    \caption{Diode effect at usual four-probe connection of the bridge: current is
applied via contacts $1$ and $3$, and voltage is measured between contacts $2$
and $4$ (see Fig.~2a) at positive and negative external magnetic
fields at $T=2$ \textrm{K}.}
    \label{fig4}
\end{figure}

It is worth mentioning that the existence of nanoholes displayed in Fig.~2b
creates an analogy between this diode behavior and the one mentioned in Ref.
\cite{Lyu21} where the conformally mapped holes were artificially introduced
in a homogeneous superconducting strip to break the inversion symmetry. In our case, the
inhomogeneously populated nanoholes exist as a consequence of
photo-projection mask imperfections which are inevitable
at the laser printing technique used for the mask preparation, 
and unexpectedly caused a positive impact. This,
however, needs more studies, since, as mentioned in \cite{Suri22,Vodolazov05,Chahid22}, the edge
imperfections can also facilitate the inversion symmetry breaking.

The main idea of the quadristor design stemmed from a hypothesis that the
underlying mechanism of this diode effect is based on vortex lattice
dynamics in the active area of the bridge. The additional leads (the pairs $%
5,6$ and $7,8$) attached to that area were added for seeding vortices in the
laminar flow of supercurrent, thus triggering major vortex pattern (turbulence)
generation. The intuitive belief that this external triggering should work
has been indeed experimentally confirmed (Fig.~5): the diode effect
via leads $1-4$ is replaced by a resistive state when the auxiliary
current through the leads $5$ and $6$ is \textit{On} 
(this numbering corresponds to that in Fig.~2a). The prediction
that there would be no latching was much less obvious. However, the removal
of the current through these auxiliary leads restored the laminar flow at
diode performance, as Fig.~5 demonstrates.
\begin{figure}
    \centering
    \includegraphics[width=\linewidth]{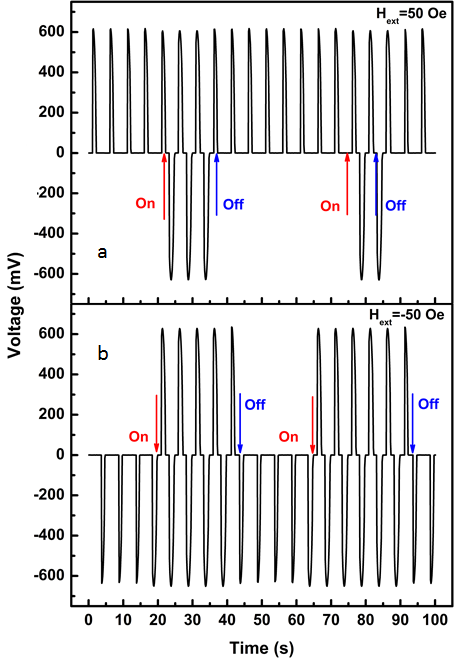}
    \caption{Performance of quadristor at application of the control current. The
transport current measurements were performed at $T=2$ \textrm{K }via main
leads $1-4$, while the control current was supplied via auxiliary leads $5,6$
or $7,8$ (see Fig.~2a for details). 
As in Fig.~4, the amplitude and frequency of the ac transport current were 9.25 mA 
(a choice dictated by results presented in Fig.~3) and 
0.2 Hz correspondingly. The control dc current was 2.5 mA (supplied by Keithley 
220 current source). The arrows indicate moments of time when
the control current was switched \textit{On} and \textit{Off}. Panels (a) and (b)
correspond to positive and negative external fields $H=\pm 50$ \textrm{Oe},
similar to Fig.~4.}
    \label{fig5}
\end{figure}
Predictably, a similar result occurs when one uses the other two auxiliary
leads ($7$ and $8$) for the control current.

The device performance resembles that of the planar
transistor-type devices suggested recently in multiple articles . 
However, in these devices, the design is based on different
principles, for example, the field-effect, which modifies the density of states,
or the injection of high-energy electrons, which generates nonequilibrium
phonon fluxes, etc. These approaches require the application of potentials in the eV range, which
is above the intrinsic characteristic energy scale of superconductors (meV). The
controlling current of our quadristor is significantly smaller than the
transport current through it, so our device can be effectively used in
circuits of superconducting microelectronics, such as logical elements,
amplifiers, etc.

In conclusion, we experimentally demonstrated a newly suggested design of
current-controlled four-terminal device which reversibly switches the
superconducting diode into a resistive state and back. The mechanism of
switching is based on nonequilibrium
kinetics of vortices in the active area of the device. Since the dioding is also
based on this mechanism and the high-frequency operation is demonstrated for it \cite{Chahid22},
there are reasons to expect that switching will also be possible at
high-frequencies. 

The aim of this research was to prove the operational
principle of the device. Importantly, the control current $(I_{cntr}=2.5$ 
\textrm{mA})  is much smaller than the amplitude of the ac transport current ($I_{tr}=9.25$ 
\textrm{mA}). At further optimization of the quadristor parameters, the gain 
$g=I_{tr}/I_{cntr}$ may be further enhanced. However, even with the currently
achieved gain, $g=3-4$, the quadristor can serve as a signal amplifier.
Moreover, the transport current through it can be controlled by many
switches (two in our current design). Thus, our development opens opportunities for various
applications in true (non-hybrid) superconducting microelectronics,
including logical units in leading-edge quantum information tasks. It also
opens novel opportunities for exploration of fundamental problems of
non-equilibrium states of superconductors and turbulent-laminar motion
interplay in superfluid liquids.

%\begin{acknowledgments}

This work is supported by the ONR grants N00014-21-1-2879 and N00014-20-1-2442.

%\end{acknowledgments}

%\appendix

%\section{Appendixes}

%\nocite{*}

%\bibliography{references.bib}

\providecommand{\noopsort}[1]{}\providecommand{\singleletter}[1]{#1}%

\end{document}